\documentclass[10pt,journal,final,twoside,twocolumn]{IEEEtran}

\usepackage[normalem]{ulem} 
\usepackage{amssymb}
\usepackage{amsmath}
\usepackage{mathtools}
\usepackage{color}
\usepackage{graphicx}
\usepackage[none]{hyphenat}
\usepackage{booktabs}
\usepackage{multirow}
\usepackage[all,poly]{xy}
\usepackage{multirow}
\usepackage{url}
\usepackage{breqn,ulem}
\usepackage{tikz}
	\usetikzlibrary{shapes,arrows}
\usepackage{pbox}
\usepackage{blindtext}
\usepackage{array}
\usepackage[ruled,vlined]{algorithm2e}
\DeclareMathOperator*{\argmin}{arg\,min}

\usepackage{url}

\bstctlcite{IEEEexample:BSTcontrol}

\begin{document}
\title{Factor analysis of dynamic PET images: \\beyond Gaussian noise}
\author{Yanna Cruz Cavalcanti,~\IEEEmembership{Member,~IEEE}, Thomas Oberlin,~\IEEEmembership{Member,~IEEE}, \\Nicolas Dobigeon,~\IEEEmembership{Senior Member,~IEEE}, C\'edric F\'evotte,~\IEEEmembership{Senior Member,~IEEE}, \\Simon Stute, Maria-Joao Ribeiro, Clovis Tauber,~\IEEEmembership{Member,~IEEE}
\thanks{Part of this work has been presented at the IEEE Int. Conf. Acoust., Speech, and Signal Processing (ICASSP), 2019 \cite{Cavalcanti2019icassp}.}
\thanks{Y. C. Cavalcanti, Th. Oberlin, N. Dobigeon and C. F\'evotte are with University of Toulouse, IRIT/INP-ENSEEIHT, CNRS, 31071 Toulouse Cedex 7, France (e-mail: \{Yanna.Cavalcanti, Thomas.Oberlin, Nicolas.Dobigeon, Cedric.Fevotte\}@irit.fr).}
    \thanks{S. Stute is with MIV, CEA, INSERM, Universit\'es Paris-Sud and Paris-Saclay, Service Hospitalier Fr\'ed\'eric Joliot, Orsay, France (e-mail:
    simon.stute@cea.fr).}
    \thanks{M.-J. Ribeiro and C. Tauber are with UMRS Inserm U930, Universit\'e de Tours, 37032 Tours, France (e-mail:
    \{maria.ribeiro, clovis.tauber\}@univ-tours.fr).}
\thanks{Part of this work has been supported by Coordena\c{c}\~ao de Aperfeiçoamento de Ensino Superior (CAPES), Brazil, and the European Research Council (ERC FACTORY-CoG-681839).}

}

\maketitle
\begin{abstract}
Factor analysis has proven to be a relevant tool for extracting tissue time-activity curves (TACs) in dynamic PET images, since it allows for an unsupervised analysis of the data. Reliable and interpretable results are possible only if considered with respect to suitable noise statistics. However, the noise in reconstructed dynamic PET images is very difficult to characterize, despite the Poissonian nature of the count-rates. Rather than explicitly modeling the noise distribution, this work proposes to study the relevance of several divergence measures to be used within a factor analysis framework. To this end, the $\beta$-divergence, widely used in other applicative domains, is considered to design the data-fitting term involved in three different factor models. The performances of the resulting algorithms are evaluated for different values of $\beta$, in a range covering Gaussian, Poissonian and Gamma-distributed noises. The results obtained on two different types of synthetic images and one real image show the interest of applying non-standard values of $\beta$ to improve factor analysis.

\end{abstract}

\vspace{-0.20cm}

\section{Introduction}
\label{sec:introduction}
\IEEEPARstart{T}{hanks} to its ability to evaluate metabolic functions in tissues from the temporal evolution of a previously injected radiotracer, dynamic positron emission tomography (PET) has become an ubiquitous analysis tool to quantify biological processes. After acquisition and reconstruction, the main time-activity curves (TACs) (herein called \textit{factors}), which represents the concentration of tracer in each tissue and blood over time, can be extracted from PET images for subsequent quantification. For this purpose, factor analysis of dynamic structures (FADS) has been intensively used \cite{Barber1980,Cavailloles1984}, further leading to FADS with nonnegative penalizations \cite{Wu1995,Sitek2000}. However, these solutions explicitly rely on the assumption that the dynamic PET noise and the model approximation errors follow Gaussian distributions.  To overcome this limitation, several works applied nonnegative matrix factorization (NMF) techniques, allowing the Kullback-Leibler (KL) divergence to be used, which is more appropriate for data corrupted by Poisson noise \cite{Lee2001a,Padilla2012,Schulz2013}. NMF with multiplicative updates is the approach generally employed since the algorithm is simple and there are less parameters to adjust than in FADS.

Nevertheless, even though the positron decay process can be described by a Poisson distribution \cite{Shepp1982}, the actual noise in reconstructed PET images is not expected to be simply described by Poisson nor Gaussian distributions. Several acquisition circumstances, such as the detector system and electronic components, as well as post-processing corrections for scatter and attenuation, significantly alter the initial Poissonian statistics of the count-rates \cite{Alpert1982,Razifar2005}. Considering the difficulties in characterizing the noise properties in PET images, many works have assumed the data to be corrupted by a Gaussian noise \cite{Fessler1994,Coxson1997,Kamasak2009}. Hybrid distributions, such as Poisson-Gaussian \cite{Slifstein2000} and Poisson-Gamma \cite{Irace2011}, have been also proposed in an attempt to take into account various phenomena occurring in the data. The work of Teymurazyan et al. \cite{Teymurazyan2012} tried to determine the statistical properties of data reconstructed by filtered-back projection (FBP) and iterative expectation maximization (EM) algorithms. While FPB reconstructed images were sufficiently described by a normal distribution, the Gamma statistics were a better fit for EM reconstructions. The recent work of Mou et al. \cite{Mou2017} further studied the Gamma behavior that can be found on PET reconstructed data.

While these works mainly put the emphasis on the noise model, the present study aims at investigating the impact of the divergence measure to be used for factor analysis of dynamic PET images. This work applies a popular and quite general loss function in NMF, namely the $\beta$-divergence \cite{Basu1998,Fevotte2011}. The $\beta$-divergence is a family of divergences parametrized by a unique scalar parameter $\beta$. In particular, it has the great advantage of generalizing conventional loss functions such as the least-square distance, KL and Itakura-Saito divergences, respectively corresponding to {additive Gaussian, Poisson and multiplicative Gamma noise}.

The current paper will empirically study the influence of $\beta$ on the factor estimation for three different methods. First, the standard $\beta$-NMF algorithm is applied. Then, an approach that includes a normalization of the factor proportions (herein called $\beta$-LMM) is used to provide factors with a physical meaning.  Finally, the $\beta$-divergence is also used to generalize the previous model introduced in \cite{Cavalcanti2018}. Simulations are conducted on two different sets of synthetic data based on realistic count-rates and one real image of a patient's brain.

This paper is organized as follows. The considered factor analysis models are described in Section \ref{sec:prob_stat}. Section \ref{sec:betadiv} presents the $\beta$-divergence as a measure of similarity. Section \ref{sec:bcd} discusses the corresponding factor analysis algorithms able to recover the factors, their corresponding proportions in each voxel and other parameters of interest. Simulation results obtained with synthetic data are reported in Section \ref{sec:synt_sim}. Experimental results on real data are provided in Section \ref{sec:real_sim}. A deeper discussion is conducted in Section \ref{sec:discuss}. Section \ref{sec:concl} concludes the paper.
\vspace{-0.15cm}
\section{Factor analysis}
\label{sec:prob_stat}
Let $\mathbf{Y}$ be an $L \times N$ observation matrix containing a 3D dynamic PET image composed of $N$ voxels acquired in $L$ time-frames. This observation matrix $\mathbf{Y}$ can be approximated by an estimated image $\mathbf{X}(\boldsymbol{\theta})$ according to a factorization model described by $P$ physically interpretable variables $\boldsymbol{\theta} = [\theta_1,\cdots,\theta_P]$, i.e.,\vspace{-0.10cm}
\begin{equation}
\mathbf{Y}\approx\mathbf{X}(\boldsymbol{\theta}).\vspace{-0.15cm}
\label{eq:observationdata}
\end{equation}
The observation image is affected by a noise whose distribution characterization is a highly challenging task, as previously explained. For this reason,  for sake of generality, the description in \eqref{eq:observationdata} makes use of an approximation symbol $\approx$ that generalizes the relation between the factor-dependent estimated image $\mathbf{X}(\boldsymbol{\theta})$ and the observed data $\mathbf{Y}$. Factor analysis can be formulated as an optimization problem which consists in estimating the parameter vector $\boldsymbol{\theta}$ assumed to belong to a set denoted $\mathcal{C}$ with possible complementary penalizations $R(\boldsymbol{\theta})$. It is mathematically described as\vspace{-0.10cm}
\begin{equation}
\hat{\boldsymbol{\theta}}\in \argmin_{\boldsymbol{\theta} \in \mathcal{C}} \Big\{\mathcal{D}( \mathbf{Y}|\mathbf{X}(\boldsymbol{\theta})) +R(\boldsymbol{\theta})\Big\}\vspace{-0.15cm}
\label{eq:optprob}
\end{equation}
where $\mathcal{D}(\cdot|\cdot)$ is a measure of dissimilarity between the observed PET image $\mathbf{Y}$ and the proposed model. The choice of this dissimilarity measure will be discussed in Section \ref{sec:betadiv}. The following paragraphs describe three different factor analysis techniques and detail particular instances of the explanatory variable $\boldsymbol{\theta}$ under this general formulation.
\vspace{-0.15cm}
\subsection{Nonnegative Matrix Factorization (NMF)}
Factorizing a latent (i.e., unobserved) matrix $\mathbf{X}\in \mathbb{R}^{{L} \times {N}}$ consists {in decomposing it} into two matrices as\vspace{-0.10cm}
\begin{equation}
\mathbf{X}=\mathbf{M}\mathbf{A},\vspace{-0.15cm}
\label{eq:nmf}
\end{equation}
where $\mathbf{M} = [\mathbf{m}_1,...,\mathbf{m}_{K}]$ is a $L \times K$ matrix of factors and $\mathbf{A} = \left[\mathbf{A}_1,\ldots,\mathbf{a}_N\right]$ is a $K \times N$ matrix containing the factor coefficients. In the dynamic PET setting, $\mathbf{M}$ is expected to contain the elementary TACs characterizing the different kinds of tissues, whereas the coefficient vector $\mathbf{a}_n$ contains their corresponding proportions in the $n$th voxel. In most applicative contexts, the number $K$ of elementary TACs is supposed to be lower than both the number of frames $L$ and the number of pixels $N$, i.e., $K \ll \min\{L,N\}$. This choice leads to a low-rank factorization of the matrix $\mathbf{X}$.

Moreover, to provide an additive and part-based description of the data, nonnegative constraints are assumed for the factors and respective proportions, resulting in the standard NMF formalism \cite{LeeNMF,lee99}\vspace{-0.10cm}
\begin{equation}
\mathbf{A} \succeq \mathbf{0}_{K,N}, \quad \mathbf{M} \succeq \mathbf{0}_{L,K},\vspace{-0.15cm}
\label{eq:constrNMF}
\end{equation}
where $\succeq$ stands for a component-wise inequality. The formulation of the corresponding NMF optimization problem has been largely considered in the literature \cite{Fevotte2011} and consists in estimating the explanatory variables $\boldsymbol{\theta} =\{\mathbf{M},\mathbf{A}\}$ subject to the constraints in \eqref{eq:constrNMF}.
\vspace{-0.15cm}
\subsection{Linear Mixing Model (LMM)}
The factorization \eqref{eq:nmf} and constraints \eqref{eq:constrNMF} that describe a typical NMF can also be envisaged under the light of the LMM widely used in the hyperspectral imagery literature \cite{Bioucas-Dias2012}. Additionally to the constraints defined in \eqref{eq:constrNMF}, to associate factors coefficients with concentrations or proportions, LMM assumes the following sum-to-one constraint\vspace{-0.10cm}
\begin{equation}
\mathbf{A}^T \mathbf{1}_{K}  = \mathbf{1}_N,\vspace{-0.15cm}
\label{eq:constrLMM}
\end{equation}
where $\mathbf{1}_N$ is the $N$-dimensional vector made of ones. The corresponding minimization problem, also widely discussed in the above-mentioned hyperspectral unmixing literature, is formulated as for {the} NMF, complemented by the additional constraint \eqref{eq:constrLMM}.
\vspace{-0.15cm}
\subsection{Specific binding linear mixing model (SLMM)}
\label{subsec:plmm}

The LMM seems to be a relevant model for dynamic PET data. Although the perfusion involved in the radiotracer diffusion is not linear, in most cases the resulting TAC is approximated by the sum of the pure TACs weighted by the factor proportions. But as discussed in \cite{Cavalcanti2018}, in high uptake regions, LMM may not provide a sufficient description of the data. Therefore, a specific binding LMM (SLMM) has been proposed to handle the variations in perfusion and labeled molecule concentration affecting the TACs related to specific binding. It describes the nonlinearity of these TACs by an additive spatially variant perturbed component that is approximated by a linear expansion over previously learned basis elements. By specifically denoting $\mathbf{M}=\left[\bar{\mathbf{m}}_1,\ldots,\mathbf{m}_K\right]$ where $\bar{\mathbf{m}}_1$ is the nominal specific binding factor, SLMM can be formulated as  \cite{Cavalcanti2018}\vspace{-0.10cm}
\begin{equation}
\mathbf{X} = \mathbf{MA} +     \underbrace{\Big[\mathbf{E}_1\mathbf{A}\cdot \mathbf{V}\mathbf{B}\Big],}_{\Delta}\vspace{-0.15cm}
\label{eq:slmm_matrix}
\end{equation}
where ``$\cdot$'' is the Hadamard point-wise product, $\mathbf{E}_{1}$ is the matrix $[\mathbf{1}_{L,1} \mathbf{0}_{L,K-1}]$, $\mathbf{V}= [\mathbf{v}_1,\ldots,\mathbf{v}_{N_v}]$ is the $L \times N_v$ matrix composed of the basis elements used to describe the variability of the specific binding factor (SBF), ($N_v \ll L$), and $\mathbf{B} = \left[\mathbf{b}_1,\ldots,\mathbf{b}_N\right]$ is the $N_v \times N$ matrix composed of internal proportions. If $\mathbf{B} = \boldsymbol{0}$, the model in  (\ref{eq:slmm_matrix}) becomes a regular linear mix, as  (\ref{eq:nmf}).

As in \cite{Cavalcanti2018}, to avoid ambiguity in the factor TACs due to their strong correlation with the variability elements, the intrinsic variability proportion matrix is constrained to be nonnegative\vspace{-0.10cm}
\begin{equation}
\mathbf{B} \succeq \boldsymbol{0}_{N_v,N}.\vspace{-0.15cm}
\label{eq:constrSLMM}
\end{equation}
Therefore, the resulting SLMM optimization problem generalizes the NMF and LMM problems
where the explanatory parameter vector is given by $\boldsymbol{\theta} = \{\mathbf{M},\mathbf{A},\mathbf{B}\}$. Under the general formalism \eqref{eq:optprob}, the set of constraints is defined by \eqref{eq:constrNMF}, \eqref{eq:constrLMM} and \eqref{eq:constrSLMM}. Moreover, as the SBF variability is only expected in the voxels belonging to the region affected by specific binding, $\mathbf{B}$ is expected to be zero outside the high-uptake region. Therefore, the spatial sparsity of the related coefficients is enforced by defining the regularizer in \eqref{eq:optprob} as\vspace{-0.10cm}
\begin{equation}
R(\boldsymbol{\theta})\triangleq \|\mathbf{B}\|_{2,1} = \sum_{n=1}^N{\|\mathbf{b}_n\|_{2}}.\vspace{-0.15cm}
\label{eq:l12}
\end{equation}

\section{Divergence measure}
\label{sec:betadiv}
When analyzing PET data, most of the works in the literature have considered the squared Euclidean distance or the Kullback-Leibler divergence as the loss function $\mathcal{D}(\cdot|\cdot)$ to be used to design the approximation model \eqref{eq:observationdata}. These choices are intrinsically related to the assumption of Gaussian and Poissonian noises, respectively. However, as previously discussed, the noise encountered in PET data is altered by several external circumstances and parameters, even though its initial count-rates is known to follow a Poisson distribution. Hence, to provide a generalization of these PET noise models, this work proposes to resort to the $\beta$-divergence as the dissimilarity measure underlying the approximation in \eqref{eq:observationdata}.

\begin{figure}[htbp]
\begin{center}
\includegraphics[width=0.6\columnwidth]{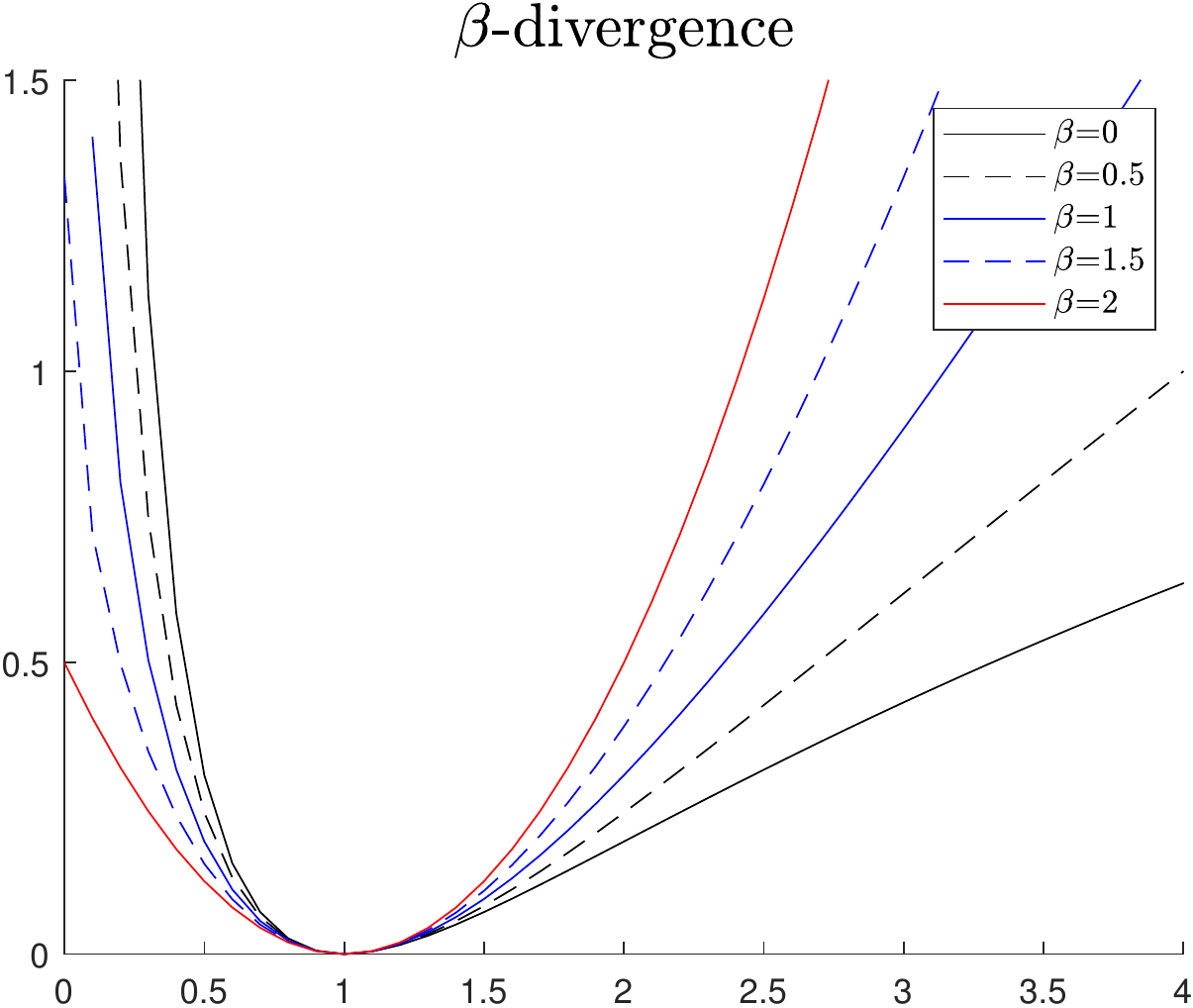}
\caption{$\beta$-divergence $d_{\beta}(y|x)$ as a function of $x$ with $y=1$ and for different values of $\beta$.}
\label{fig:betadiv}
\end{center}
\end{figure}

The $\beta$-divergence first appeared in the works of Basu et al. \cite{Basu1998} and Eguchi and Kano \cite{Eguchi2001}. Since then, it has been intensively used, with noticeable successes in the audio literature for music transcription and separation \cite{OGRADY2008,Fitzgerald2009,Fevotte2009}. More precisely, the $\beta$-divergence between two matrices $\mathbf{Y}$ and $\mathbf{X}$ follows the component-wise separability property\vspace{-0.10cm}
\begin{equation}
 \mathcal{D}_{\beta}(\mathbf{Y}|\mathbf{X}) = \sum_{\ell=1}^L \sum_{n=1}^N {d}_{\beta}(y_{\ell,n}|x_{\ell,n})\vspace{-0.15cm}
\end{equation}
and is defined for {$\beta \in \mathbb{R}$ as}
{
\begin{multline}
{d}_{\beta}(y|x) = \\
                \begin{cases}
                \frac{1}{\beta(\beta-1)}(y^{\beta}+(\beta-1)x^{\beta}-\beta yx^{\beta-1}) & \beta \in \mathbb{R}\backslash \{0,1 \}\\
                y\log{\frac{y}{x}}-y+x & \beta=1,\\
                \frac{y}{x}-\log{\frac{y}{x}}-1 & \beta=0.\\
                \end{cases}\vspace{-0.15cm}
\label{eq:betalim}
\end{multline}}
{The limit cases $\beta=1,0$ correspond to the KL and IS divergences, respectively, while $\beta=2$ coincides with the squared Euclidean distance. As an illustration}, Fig. \ref{fig:betadiv} compares the loss functions ${d}(y=1|x)$ as functions of $x$ for various values of $\beta$. For a comprehensive discussion of the  $\beta$-divergence, the interested {readers are} invited to consult \cite{Cichocki2010}.

Among its interesting properties, the $\beta$-divergence can be related to a wide family of distributions, namely the Tweedie distributions, via its corresponding density $p(y|x)$ following\vspace{-0.10cm}
\begin{equation}
-\log p(y|x) = \varphi^{-1} d_{\beta}(y|x) + \text{const.}\vspace{-0.15cm}
\end{equation}
where $\varphi$ is a so-called dispersion parameter \cite{Tan2013}. In particular, the Tweedie distributions encompass a large class of popular distributions, including the Gaussian, Poissonian and Gamma distributions. In other words, choosing the $\beta$-divergence as the loss function in \eqref{eq:optprob} allows the approximation  \eqref{eq:observationdata} to stand for a wide range of noise models. For instance, the $\beta$-divergence in the special cases $\beta=2,1,0$ is related to additive Gaussian, Poisson and multiplicative Gamma observation noises \cite{Fevotte2011}. As a consequence, thanks to its genericity, the $\beta$-divergence seems to {be a relevant} tool to conduct factor analysis when the PET noise is difficult to be characterized.

\section{Block-coordinate descent algorithm}
\label{sec:bcd}
The non-convex minimization problem stated in (\ref{eq:optprob}) is solved through a block-coordinate descent (BCD) algorithm. For each factor analysis model discussed in Section \ref{sec:prob_stat}, the corresponding algorithm iteratively updates a latent variable ${\theta}_i$ while all the others are kept fixed, allowing for convergence towards a local solution. The definition of these blocks naturally arises according to  the considered latent factor model. The method detailed hereafter leads to multiplicative update rules, i.e., consists in multiplying the current variable values by nonnegative terms, thus preserving the nonnegativity constraint along the iterations. To avoid undesirable solutions, given the non-convexity of the problem, the algorithms require proper initialization.

The algorithm and corresponding updates used for $\beta$-NMF has been introduced in \cite{Fevotte2011}. Therefore, the present paper derives only the algorithm associated with the SLMM model, that turns into LMM when fixing $\mathbf{B}=\mathbf{0}$. The updates are derived following the strategy proposed in \cite{Fevotte2015}, while some heuristic rules are inspired by \cite{Fevotte2009}. The principles of these updates are briefly recalled in paragraph \ref{subsec:mm_algo} and particularly instantiated for the considered SLMM-based factor model in paragraphs \ref{subsec:mm_algo_M}--\ref{subsec:mm_algo_B}. For conciseness, we only present derivations for $\beta \in [1,2]$ (the interval where $d_\beta(x|y)$ is convex with respect to $y$)  but they can be easily generalized for other values using the methodology described in \cite{Fevotte2011,yang11,Fevotte2015}. The resulting algorithmic procedure is summarized in Algo. \ref{algo:globalslmm} where all multiplications (identified by the $\cdot$ symbol), divisions and exponentiations are entry-wise operations, $\mathbf{1}_{K,L}$ denote a $K\times L$ matrix of ones and $\boldsymbol{\Gamma}_{\mathbf{B}} \triangleq \mathrm{diag}[\|\mathbf{b}_1\|_1,\cdots,\|\mathbf{b}_1\|_N]^{-1}$. { Note that, although this algorithmic resolution differs from the one initially proposed in \cite{Cavalcanti2018}, the final results obtained by setting $\beta=2$ are very similar for the same parameter values.}

    \LinesNumbered
    \begin{algorithm}
    \DontPrintSemicolon
    \KwData{$\mathbf{Y}$}
    \KwIn{$\mathbf{A}^{0}$, $\mathbf{M}^{0}$, $\mathbf{B}^{0}$, $\lambda$}
    $k \leftarrow 0$\;
    $\tilde{\mathbf{Y}} \leftarrow \mathbf{M}^{0}\mathbf{A}^{0} + \Big[\mathbf{E}_1\mathbf{A}^{0}\cdot \mathbf{V}\mathbf{B}^{0}\Big]$\;
    \While{stopping criterion not satisfied}{
    \% Update variability matrix \newline
        \label{algostep:B} $\mathbf{B}^{k+1} \leftarrow
        \mathbf{B}^{k}\cdot\Bigg[\frac{\mathbf{1}_{N_v}^T\mathbf{A}_{1,:}\cdot(\mathbf{V}^T(\mathbf{Y}\cdot\tilde{\mathbf{X}}^{\beta-2}))}{\mathbf{1}_{N_v}^T\mathbf{A}_{1,:}\cdot(\mathbf{V}^T\tilde{\mathbf{X}}^{\beta-1})+\lambda\mathbf{B}^k\boldsymbol{\Gamma}_{\mathbf{B}}}\Bigg]^{\frac{1}{3-\beta}}$ \;
    $\tilde{\mathbf{X}} \leftarrow \mathbf{M}^{k}\mathbf{A}^{k} + \Big[\mathbf{E}_1\mathbf{A}^{k}\cdot \mathbf{V}\mathbf{B}^{k+1}\Big]$\;
    \% Update factor TACs \newline
    \label{algostep:M} $\mathbf{M}^{k+1}_{2:K} \leftarrow  \mathbf{M}^{k}_{2:K} \Bigg[\frac{(\mathbf{Y}\cdot\tilde{\mathbf{X}}^{\beta-2})\mathbf{A}^{T}_{2:K}}{\tilde{\mathbf{X}}^{\beta-1}\mathbf{A}^{T}_{2:K}} \Bigg]$\;
        $\tilde{\mathbf{X}} \leftarrow \mathbf{M}^{k+1}\mathbf{A}^{k} + \Big[\mathbf{E}_1\mathbf{A}^{k}\cdot \mathbf{V}\mathbf{B}^{k+1}\Big]$\;

    \% Update SBF factor proportion \newline
    \label{algostep:A1} ${\mathbf{A}^{k+1}_{1} \leftarrow \mathbf{A}^{k}_{1}\cdot\Bigg[\frac{\mathbf{1}_{L}^T((\mathbf{M}_{1}\mathbf{1}_{N}^T+\mathbf{VB})\cdot(\mathbf{Y}\cdot\tilde{\mathbf{X}}^{\beta-2})+\mathbf{\tilde{X}}^{\beta})}{\mathbf{1}_{L}^T((\mathbf{M}_{1}\mathbf{1}_{N}^T+\mathbf{VB})\cdot\tilde{\mathbf{X}}^{\beta-1}+\mathbf{Y}\cdot\tilde{\mathbf{X}}^{\beta-1})} \Bigg]}$\;

    \% Update other factor proportions \newline
    \label{algostep:Ak} $\mathbf{A}^{k+1}_{2:K} \leftarrow \mathbf{A}^{k}_{2:K}\cdot\Bigg[\frac{\mathbf{M}_{2:K}^{T}(\mathbf{Y}\cdot\tilde{\mathbf{X}}^{\beta-2})+\mathbf{1}_{K-1,L}\tilde{\mathbf{X}}^{\beta}}{\mathbf{M}_{2:K}^T\tilde{\mathbf{X}}^{\beta-1}+\mathbf{1}_{K-1,L}(\mathbf{Y}\cdot\tilde{\mathbf{X}}^{\beta-1})} \Bigg]$\;

    $k \leftarrow k+1$\;
    $\tilde{\mathbf{X}} \leftarrow \mathbf{M}^{k}\mathbf{A}^{k} + \Big[\mathbf{E}_1\mathbf{A}^{k}\cdot \mathbf{V}\mathbf{B}^{k}\Big]$\;
    }
    $\mathbf{A} \leftarrow \mathbf{A}^{k}$\;
    $\mathbf{M} \leftarrow \mathbf{M}^{k}$\;
    $\mathbf{B} \leftarrow \mathbf{B}^{k}$\;
    \KwResult{$\mathbf{A}$, $\mathbf{M}$, $\mathbf{B}$}
    \caption{$\beta$-SLMM unmixing \label{algo:globalslmm}}
    \end{algorithm}

\subsection{Majorization-minimization and multiplicative algorithms} \label{subsec:mm_algo}
Our methodology relies on majorization-minimization (MM) and multiplicative algorithms that are common to many NMF settings. Majorization-minimization (MM) algorithms consist in finding a surrogate function that majorizes the original objective function and then computing its minimum. The algorithm iteratively updates each variable $\theta_i$ given all the other variables $\boldsymbol{\theta}_{j\neq i}$. Hence, the subproblems to be solved can be written
\begin{equation}\label{eq:min_J}
\min_{\theta_i} \mathcal{J}(\theta_i) = \mathcal{D}(\mathbf{Y}|\mathbf{X}(\boldsymbol{\theta}))+R(\theta_i) \text{ s.t. } \theta_i \in \mathcal{C}.
\end{equation}
By denoting  $\tilde{\theta}_i$ the state of the latent variable $\theta_i$ at the current iteration, we first define an auxiliary function $G(\theta_i|\tilde{\theta}_i)$ that majorizes $\mathcal{J}(\theta_i) $, i.e., $G(\theta_i|\tilde{\theta}_i) \geq \mathcal{J}(\theta_i)$, and is tight at $\tilde{\theta}_i$, i.e. $G(\tilde{\theta}_i|\tilde{\theta}_i) = \mathcal{J}(\tilde{\theta}_i)$. The optimization problem \eqref{eq:min_J} is then replaced by the minimization of the auxiliary function. {In many NMF problems, canceling the auxiliary function gradient} leads to multiplicative updates of the form\vspace{-0.10cm}
\begin{equation}
\theta_i = \tilde{\theta}_i\left[\frac{N(\tilde{\theta}_i)}{D({\tilde{\theta}_i})}\right]^{{\gamma}}
\label{eq:MMupdates}\vspace{-0.15cm}
\end{equation}
where the functions $N(\cdot)$, $D(\cdot)$ and the scalar exponent $\gamma$ are problem-dependent.

A heuristic alternative to this algorithm is described in \cite{Fevotte2009}. It consists in decomposing the gradient {of the objective function $\cal{J}$} with respect to (w.r.t.) the variable $\tilde{\theta}_i$ as the difference between two nonnegative functions, such that\vspace{-0.10cm}
\begin{equation}
\nabla_{\theta_i} \mathcal{J}(\tilde{\theta}_i) = \nabla_{\theta_i}^+ \mathcal{J}(\tilde{\theta}_i)-\nabla_{\theta_i}^- \mathcal{J}(\tilde{\theta}_i)\vspace{-0.15cm}
\end{equation}
and using
\eqref{eq:MMupdates} with\vspace{-0.10cm}
\begin{eqnarray}
N(\tilde{\theta}_i) &=& {\nabla_{\tilde{\theta}_i}^- \mathcal{J}(\tilde{\theta}_i)} \label{eq:heurialgo_1},\\
D(\tilde{\theta}_i) &=& {\nabla_{\tilde{\theta}_i}^+ \mathcal{J}(\tilde{\theta}_i)}\label{eq:heurialgo_2}.\vspace{-0.15cm}
\end{eqnarray}
The heuristic and MM algorithms coincide in many well-known cases \cite{Kompass2007,Fevotte2011,yang11}. MM guarantees monotonic decrease of the objective function at every iteration. This is not guaranteed by the heuristic alternative, but is often observed in practice \cite{yang11}. Note that monotonic decrease of the objective function does not automatically implies convergence of the parameter iterates, though this is also typically observed in practice.%

\subsection{Update of the factor TACs $\mathbf{M}$}\label{subsec:mm_algo_M}
According to the optimization framework described above, given the current values ${\mathbf{A}}$ and ${\mathbf{B}}$ of the abundance matrix  and the internal proportions, respectively, updating the factor matrix $\mathbf{M}$ can be formulated as the minimization subproblem\vspace{-0.10cm}
\begin{equation}
\min_{\mathbf{M}} \mathcal{J}(\mathbf{M}) = \mathcal{D}( \mathbf{Y}|\mathbf{M}{\mathbf{A}}+\Delta) \text{ s.t. }\mathbf{M} \succeq \mathbf{0}_{L,K},\vspace{-0.15cm}
\end{equation}
{with $\Delta=\mathbf{E}_1\mathbf{A}\cdot \mathbf{V}\mathbf{B}$}. Following \cite{Fevotte2015}, when $\beta \in [1,2]$, the objective function $\mathcal{J}(\mathbf{M})$ can be simply majorized using Jensen's inequality:\vspace{-0.10cm}
\begin{equation}
\hspace{-0.5cm}\mathcal{J}(\mathbf{M})\leq \underbrace{\sum_{lnk} \bigg[\frac{\tilde{m}_{lk}a_{kn}}{\tilde{x}_{ln}}d(y_{ln}|\frac{\tilde{x}_{ln}m_{lk}}{\tilde{m}_{lk}})+\frac{\delta_{ln}}{\tilde{x}_{ln}}d(y_{ln}|\tilde{x}_{ln})\bigg]}_{G(\mathbf{M}|\tilde{\mathbf{M}})}\vspace{-0.15cm}
\end{equation}
where {$\tilde{x}_{ln}=\sum_k\tilde{m}_{lk}a_{kn}+\delta_{ln}$} is the current state of the model-based reconstructed data. The auxiliary function $G(\mathbf{M}|\tilde{\mathbf{M}})$ essentially majorizes the divergence of the sum by the sum of the divergences, allowing the optimization of $\mathbf{M}$ to be conducted element-by-element.
The gradient w.r.t. the element $m_{lk}$ writes\vspace{-0.15cm}
\begin{equation}
\begin{aligned}
\nabla_{m_{lk}} G(\mathbf{M}|\tilde{\mathbf{M}}) &= \sum_n{a_{kn}\tilde{x}_{ln}}^{\beta-1}\bigg(\frac{m_{lk}}{\tilde{m}_{lk}}\bigg)^{\beta-1}\\\vspace{-0.15cm}
&-\sum_n{a_{kn}y_{ln}\tilde{x}_{ln}^{\beta-2}\bigg(\frac{m_{lk}}{\tilde{m}_{lk}}\bigg)^{\beta-2}}.\vspace{-0.15cm}
\end{aligned}
\end{equation}
Thus, minimizing $ G(\mathbf{M}|\tilde{\mathbf{M}})$ w.r.t. $\mathbf{M}$ leads to the following element-wise multiplicative update
\begin{equation}
m_{lk} = \tilde{m}_{lk}\Bigg[\frac{\sum_n{a_{kn}y_{ln}\tilde{x}_{ln}^{\beta-2}}}{\sum_n{a_{kn}\tilde{x}_{ln}^{\beta-1}}}\Bigg]^{\gamma(\beta)}.\vspace{-0.15cm}
\end{equation}
where $\gamma(\beta) = 1$ when $\beta \in [1,2]$. More generally, it can be shown that the update is still valid for $\beta \not \in [1,2]$, with $\gamma(\beta) = \frac{1}{2- \beta}$ for $\beta<1$ and $\gamma(\beta) = \frac{1}{\beta-1}$ for $\beta>2$ \cite{Fevotte2015}.

\subsection{Update of the factor proportions $\mathbf{A}$}\label{subsec:mm_algo_A}
Given the current values ${\mathbf{M}}$ and ${\mathbf{B}}$ of the factor matrix and internal propositions, the update rule for $\mathbf{A}$ is obtained by solving\vspace{-0.10cm}
\begin{equation}
\begin{aligned}
\min_{\mathbf{A}} \mathcal{J}(\mathbf{A}) &= \mathcal{D}( \mathbf{Y}|{\mathbf{M}}\mathbf{A}+\Big[\mathbf{E}_1\mathbf{A}\cdot {\mathbf{W}})\Big]) \\
&\text{ s.t. }\mathbf{A} \succeq \mathbf{0}_{K,N}\text{, } \mathbf{A}^T\mathbf{1}_{K} = \mathbf{1}_{N},\vspace{-0.15cm}
\end{aligned}
\end{equation}
with $\mathbf{W}=\mathbf{V}\mathbf{B}$. Constructing a MM algorithm that enforces the sum-to-one constraint is not straightforward and we instead resort to the method described in \cite{Eggert2004,Fevotte2015}, which relies on a change of variable. More precisely, by introducing an auxiliary matrix $\mathbf{U}$, the components $a_{kn}$ of the factor proportion matrix $\mathbf{A}$ can be rewritten as\vspace{-0.10cm}
\begin{equation}
a_{kn} = \frac{u_{kn}}{\sum_ku_{kn}},\vspace{-0.15cm}
\end{equation}
which explicitly ensures the sum-to-one constraint \eqref{eq:constrLMM}.
The new optimization problem is then\vspace{-0.10cm}
\begin{equation}
\min_{\mathbf{U}} \mathcal{J}(\mathbf{U}) \text{ s.t. }\mathbf{U} \succeq \mathbf{0}_{K,N},\vspace{-0.15cm}
\end{equation}
with\vspace{-0.10cm}
\begin{align}
\mathcal{J}(\mathbf{U})&
= \mathcal{D}( \mathbf{Y}|{\mathbf{M}}\bigg[\frac{\mathbf{u}_1}{\|\mathbf{u}_1\|_1},\cdots,\frac{\mathbf{u}_N}{\|\mathbf{u}_N\|_1}\bigg]\nonumber\\
& \qquad +\Big[\mathbf{E}_1\bigg[\frac{u_{11}}{\|\mathbf{u}_1\|_1},\cdots,\frac{u_{1N}}{\|\mathbf{u}_N\|_1}\bigg]\cdot {\mathbf{W}})\Big]) \nonumber\\
&=\sum_{ln}d(y_{ln}|\sum_k {m}_{lk}\bigg[\frac{u_{kn}}{\|\mathbf{u}_n\|_1}\bigg]+\bigg[\frac{u_{1n}}{\|\mathbf{u}_n\|_1}\bigg]{w}_{ln}).\vspace{-0.15cm}
\end{align}
Unfortunately, constructing a MM algorithm for $\mathbf{U}$ is not straightforward. As such, we resort to the heuristic alternative described in paragraph \ref{subsec:mm_algo}. Denoting $\tilde{x}_{ln}=\sum_{k\neq 1} {m}_{lk}\tilde{a}_{kn}+\tilde{a}_{1n}{w}_{ln}$, this leads to the following multiplicative update:
\begin{equation*}
u_{kn} = \tilde{u}_{kn} \, r_{kn}\vspace{-0.15cm}
\end{equation*}
where \vspace{-0.10cm}
\begin{equation*}
r_{kn} = \left\{
  \begin{array}{ll}
     \frac{\sum_l\big(\tilde{x}_{ln}^{\beta}+({m}_{l1}+{w}_{ln})\tilde{x}_{ln}^{\beta-2}y_{ln}\big)}{\sum_l\big(({m}_{l1}+{w}_{ln})\tilde{x}_{ln}^{\beta-1}+y_{ln}\tilde{x}_{ln}^{\beta-1}\big)}, & \hbox{if $k=1$;} \\
     \frac{\sum_l\big(\tilde{x}_{ln}^{\beta}+{m}_{lk}y_{ln}\tilde{x}_{ln}^{\beta-2}\big)}{\sum_l\big({m}_{lk}\tilde{x}_{ln}^{\beta-1}+y_{ln}\tilde{x}_{ln}^{\beta-1}\big)}, & \hbox{otherwise.}
  \end{array}
\right.\vspace{-0.15cm}
\end{equation*}

\subsection{Update of the internal variability $\mathbf{B}$}\label{subsec:mm_algo_B}
Given the current states ${\mathbf{M}}$ and ${\mathbf{A}}$ of the factor matrix and factor proportions, respectively, updating $\mathbf{B}$ consists in solving\vspace{-0.10cm}
\begin{equation}
\begin{aligned}
\min_{\mathbf{B}} \mathcal{J}(\mathbf{B}) = \mathcal{D}( \mathbf{Y}|{\mathbf{M}}{\mathbf{A}}&+\Big[\mathbf{E}_1{\mathbf{A}}\cdot \mathbf{VB})\Big])+\lambda \|\mathbf{B}\|_{2,1} \\
&\text{ s.t. }\mathbf{B} \succeq \mathbf{0}_{N_v,N},\vspace{-0.15cm}
\end{aligned}
\end{equation}
where the parameter $\lambda$ controls the trade-off between the data-fitting term and the spatial sparsity-inducing regularization $\|\mathbf{B}\|_{2,1}$. Denoting by $\mathbf{\tilde{B}}$ the current state of $\mathbf{B}$, the model-based reconstructed data using the current estimates is now defined by $\tilde{x}_{ln}=s_{ln}+\sum_{i}a_{1n}v_{li}\tilde{b}_{in}$ with $s_{ln}=\sum_km_{lk}a_{kn}$.

Assuming $\beta \in [1,2]$, Jensen's inequalities allows the data fitting term to be majorized as\vspace{-0.15cm}
\begin{align}
\mathcal{D}( \mathbf{Y}&|\mathbf{S}+[\mathbf{E}_1\mathbf{A}\cdot \mathbf{VB}])\nonumber\\
&\leq \underbrace{\sum_{ln}\bigg[ \frac{s_{ln}}{\tilde{x}_{ln}}d(y_{ln}|\tilde{x}_{ln})+\sum_{i}\frac{a_{1n}v_{li}\tilde{b}_{in}}{\tilde{x}_{ln}}d(y_{ln}|\frac{\tilde{x}_{ln}b_{in}}{\tilde{b}_{in}})\bigg]}_{F(\mathbf{B}|\tilde{\mathbf{B}})}.\nonumber\vspace{-0.15cm}
\end{align}
The auxiliary function associated with $\mathcal{J}(\mathbf{B})$ can be decomposed as $G(\mathbf{B}|\tilde{\mathbf{B}})=F(\mathbf{B}|\tilde{\mathbf{B}})+\lambda \|\mathbf{B}\|_{2,1}$. However, minimizing this auxiliary function w.r.t. $\mathbf{B}$ is not straightforward. Following \cite{Fevotte2015}, the regularization $\|\mathbf{B}\|_{2,1}$ is majorized itself by a tangent inequality, thanks to the concavity of the square-root function:
\vspace{-0.10cm}
\begin{equation}
\|\mathbf{B}\|_{2,1}\leq \underbrace{\frac{1}{2}\sum_n \bigg(\frac{\|\mathbf{b}_n\|_2^2}{\|\tilde{\mathbf{b}}_n\|_2}+\|\tilde{\mathbf{b}}_n\|_2\bigg)}_{H(\mathbf{B}|\tilde{\mathbf{B}})}.\vspace{-0.15cm}
\label{eq:normmajor}
\end{equation}
Unfortunately, the resulting auxiliary function is not yet amenable to optimization and our approach again closely follows \cite{yang11,Fevotte2015}. The leading monomial of $F(\mathbf{B}|\tilde{\mathbf{B}})$ (of degree lower than 2 when $\beta \in [1,2]$) must be majorized by a quadratic term, matching the quadratic upper bound of the penalty function. After canceling the gradient of the resulting auxiliary function, this leads to the multiplicative update
\begin{equation}
b_{in} = \tilde{b}_{in}\bigg(\frac{a_{1n}\sum_{l}v_{li}y_{ln}\tilde{x}_{ln}^{\beta-2}}{a_{1n}\sum_{l}v_{li}\tilde{x}_{ln}^{\beta-1}+\lambda\frac{\tilde{b}_{in}}{\|\tilde{\mathbf{b}}_n\|_2}}\bigg)^{\xi(\beta)},\vspace{-0.15cm}
\end{equation}
where $\xi(\beta) = \frac{1}{3-\beta}$ when $\beta \in [1,2]$. Again, the update can be generalized to $\beta \in \mathbb{ R}$ following \cite{Fevotte2015} and references therein. Experiments showed that dropping the exponent $\xi(\beta)$ still results in a valid algorithm while accelerating convergence.

\section{Experiments with synthetic data}
\label{sec:synt_sim}
\subsection{Synthetic data generation}
Simulations have been conducted on synthetic images with realistic count-rate properties \cite{Stute2015}. These images have been generated from the Zubal high resolution numerical phantom \cite{zubal1994computerized} with values derived from real PET images acquired with the Siemens HRRT using the $^{11}$C-PE2I radioligand. The original phantom data is of size $256 \times 256\times 128$ with a voxel size of $1.1 \times 1.1\times 1.4$ mm$^3$ , and was acquired over $L=20$ frames of durations that range from $1$ to $5$ minutes for a $60$ minutes total acquisition.

\subsubsection{Phantom I generation}
A clinical PET image with $^{11}$C-PE2I of a healthy control subject has been segmented into regions-of-interest using a corresponding magnetic resonance image. Then averaged TACs of each region have been extracted and set as the TAC of voxels in the corresponding phantom region. It is worth noting that this supervised segmentation neglects any labeled molecule concentration differences due to possible variability in the specific binding region. Thus, it describes each entire segmented region by a single averaged TAC. This phantom, referred to as Phantom I, has been used to evaluate the reconstruction error for different values of $\beta$.

\subsubsection{Phantom II generation}
To evaluate the impact of $\beta$ on the factor analysis, a second synthetic phantom, referred to as Phantom II, has been also created as follows. Phantom I has been unmixed with the N-FINDR \cite{Winter1999spie} to extract $K=4$ factors \cite{Yaqub2012} that correspond to tissues of the brain: specific gray matter, blood or veins, white matter and non-specific gray matter. The corresponding ground truth factor proportions have been subsequently set as those estimated by SUnSAL \cite{Bioucas2010}. Then, the SBF as well as the variability dictionary have been generated from a compartment model \cite{Phelps1986}, while the internal variability have been generated by dividing the region concerned by specific binding into 4 subregions with different mean variabilities. Phantom II is finally obtained by mixing these ground truth components according to SLMM in \eqref{eq:slmm_matrix}.

\subsubsection{Dynamic PET image simulation}
\label{subsec:synth_sim_gen}
The generation process that takes realistic count rates properties into consideration is detailed in \cite{Stute2015}. To summarize, activity concentration images are first computed from the input phantom and TACs, applying the decay of the positron emitter with respect to the provided time frames. To mimic the partial volume effect, a stationary 4mm FWHM isotropic 3D Gaussian point spread function {(PSF)} is applied, followed by a down-sampling to a 128 x 128 x 64 image matrix of 2.2 x 2.2 x 2.8 mm$^3$ voxels. Data is then projected with respect to real crystal positions of the Siemens Biograph TruePoint TrueV scanner, taking attenuation into account. A scatter distribution is computed from a radial convolution of this signal. A random distribution is computed from a fan-sum of the true-plus-scatter signal. Realistic scatter and random fractions are then used to scale all distributions and compute the prompt sinograms. Finally, Poisson noise is applied based on a realistic total number of counts for the complete acquisition. {Data were reconstructed using the standard ordered-subset expectation maximization (OSEM) algorithm (16 subsets) including a 4mm FWHM 3D Gaussian PSF modeling as in the simulation.} Two images, referred to as \emph{6it} and \emph{50it}, are considered for the analysis: the 6th iteration without post-smoothing, and the 50th iteration post-smoothed with {a 4mm FWHM 3D Gaussian kernel} \cite{stute2013}. A set of 64 independent samples of each phantom were generated to assess consistent statistical performance.

\subsection{Compared methods}
\subsubsection{Phantom I}
The main objective when using Phantom I is to evaluate the influence of $\beta$ on the factor modeling (i.e., by evaluating the reconstruction error) for images reconstructed with $6$ and $50$ iterations. It also provides a relevant   comparison between $\beta$-NMF  and the {more constrained solution recovered by $\beta$-LMM}. Within this experimental setup, $\beta$ ranges from $0$ to $2.4$ with a step size of $0.2$. Factor TACs are initialized by vertex component analysis (VCA) \cite{Nascimento2005}, while the factor proportions are initialized {either by SUnSAL or }randomly, depending on the considered setting (see paragraph \ref{subsec:results_phantomI}). The algorithmic stopping criterion, relating the past and current states of the objective function $\mathcal{J}$, is defined as $\frac{\mathcal{J}^{(i-1)}-\mathcal{J}^{(i)}}{\mathcal{J}^{(i-1)}} < \varepsilon$, where the values of $\varepsilon$ are reported in Table \ref{table:param}.

\subsubsection{Phantom II}
For the sake of comparison, Phantom II will be analyzed with both the $\beta$-SLMM algorithm and its simpler version, $\beta$-LMM, which does not take variability into account. {The corresponding algorithms are applied for $\beta\in\left\{0,1,2\right\}$. Since Phantom II exhibits a high variability in the tissue corresponding to the SBF, the pure-pixel assumption considered in VCA may not be enough to capture the complexity of the mixture. For this reason, factor TACs have been initialized with K-means, which is more robust to outliers.} Factor proportions have been initialized either with SUnSAL either randomly, depending on the considered setting (see paragraph \ref{subsec:results_phantomII}). The variability matrix $\mathbf{B}$ is randomly initialized on both settings. The values for $\varepsilon$ in Table \ref{table:param} are also valid in this setting.

\subsection{Performance measures}
\subsubsection{Phantom I}
In the first round of experiments, the reconstruction error is computed in terms of {peak signal-to-noise ratio (PSNR)}
\begin{equation}
\label{eqref:PSNR}
\mathrm{PSNR}(\hat{\mathbf{X}}) = 10\log_{10}{\frac{\max(\mathbf{X}^*)^2}{ \| \hat{\mathbf{X}}-\mathbf{X}^*\|_F^2}}
\end{equation}
where $\max(\mathbf{X}^*)$ is the maximum value of the ground-truth latent image $\mathbf{X}^*$ and $\hat{\mathbf{X}} \triangleq \mathbf{X}(\hat{\boldsymbol{\theta}})$ is the image recovered according to the considered factor model \eqref{eq:observationdata} with the estimated latent variables $\hat{\boldsymbol{\theta}}$.

\subsubsection{Phantom II}
In addition to the PSNR, performances on Phantom II have been evaluated w.r.t. each latent variable by computing the normalized mean square error (NMSE):
\begin{equation}
 \mathrm{NMSE}(\hat{\theta}_i)=\frac{ \| \hat{\theta}_i-\theta_i^*\|_F^2}{ \|\theta_i^*\|_F^2},
\end{equation}
where $\theta_i^*$ and $\hat{\theta}_i$ are the actual and estimated latent variables, respectively. In particular, the NMSE has been computed for the following variables: the high-uptake factor proportions $\mathbf{A}_1$, the remaining factor proportions $\mathbf{A}_{2:K}$, the SBF TAC $\tilde{\mathbf{M}}_{1}$, the non-specific factor TACs $\mathbf{M}_{2:K}$ and finally, when considering $\beta$-SLMM, the internal variability $\mathbf{B}$.

\subsection{Results on Phantom I}
\label{subsec:results_phantomI}
In the first round of simulations, $\beta$-NMF and $\beta$-LMM algorithms are evaluated in terms of the reconstruction error \eqref{eqref:PSNR} for several values of $\beta$. Two cases are considered. The first one considers that the factor TACs previously estimated by VCA are fixed. Thus, the algorithm described in Section \ref{sec:bcd} updates only the factor proportions, within a convex optimization setting. In this case, the factor proportions have been randomly initialized. Within the second and non-convex setting, the algorithm estimates both factor TACs and proportions where the factor proportions have been initialized using SUnSAL. Note that complementary results are reported in the companion report \cite{Cavalcanti2018TR}.

\subsubsection{$\beta$-NMF results}

\begin{figure}[htbp]
\begin{center}
\includegraphics[width=\columnwidth]{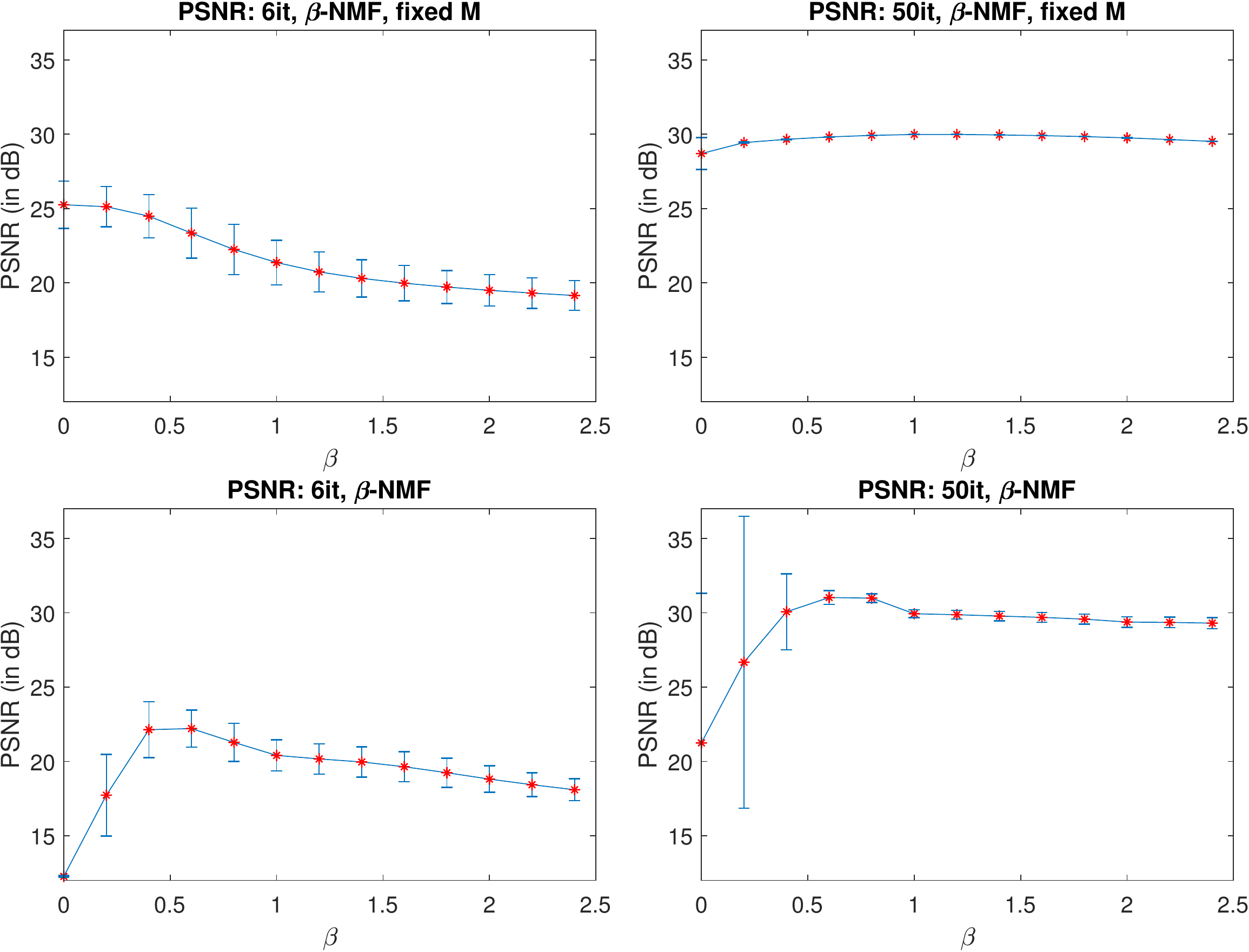}
\caption{PSNR mean and standard deviation obtained on the 6it (left) and 50it (right) images after factorization with $\beta$-NMF with fixed (top) and estimated (bottom) factor TACs over 64 samples.}
\label{fig:psnr_nmf}
\end{center}
\end{figure}

Figure \ref{fig:psnr_nmf} shows the PSNR mean and corresponding standard deviation obtained on the 6it and 50it images when analyzed with $\beta$-NMF. The first line corresponds to the the convex estimation setting (i.e., fixed factor TACs) while the non-convex framework (i.e., estimated factor TACs) is reported in the second line. The 6it images show higher PSNRs for the values of $\beta \in [0,0.6]$ in both convex and non-convex settings. This result indicates a residual noise that is rather between Gamma and Poisson distributed, which is consistent with previous studies from the literature \cite{Teymurazyan2012,Mou2017}. The best performance $\mathrm{PSNR}=25$dB with fixed $\mathbf{M}$ is reached for $\beta=0$, which significantly outperforms the result obtained with the Euclidean divergence $\beta=2$ commonly adopted  in the literature. Within a non-convex optimization setting, when estimating both factor TACs and proportions, the maximum $\mathrm{PSNR}=22.2$dB is obtained  for $\beta=0.6$ , followed by $\beta=0.4$. In this case, the difference between the greater and smaller PSNRs is of almost 3.5 dB. As non-convex optimization problems are highly sensitive to the initialization, the convex frameworks shows a better mean performance for all values of $\beta$, as well as less variance among the different realizations.

The reconstruction of the 50it images is clearly less sensitive to the choice of the divergence. Yet, values $\beta=1$  and $\beta=0.5$  in the convex and non-convex settings, respectively, increase the reconstruction PSNR by about $1$dB. This is consistent with prior knowledge about the noise statistics: whereas  the nature of noise in the 50it image is still Poissonian, its power is very low due to a higher level of filtering.

\subsubsection{$\beta$-LMM results}

\begin{figure}[htbp]
\begin{center}
\includegraphics[width=\columnwidth]{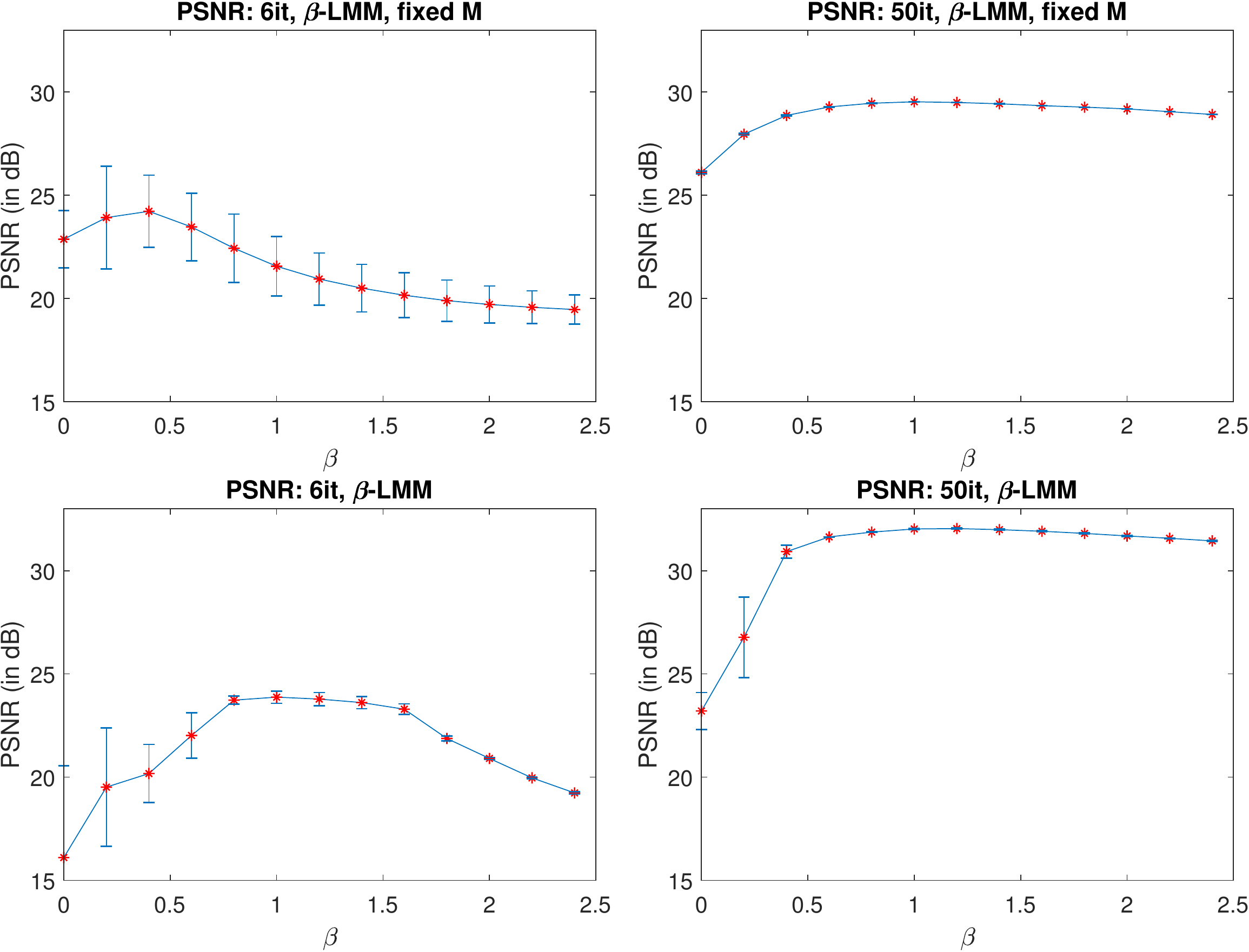}
\caption{PSNR mean and standard deviation obtained on the 6it (left) and 50it (right) images after factorization with $\beta$-LMM with fixed (top) and estimated (bottom) factor TACs over 64 samples.}
\label{fig:psnr_lmm}
\end{center}
\end{figure}

Figure \ref{fig:psnr_lmm} shows the PSNR mean and standard deviation after factorization with $\beta$-LMM with fixed (top) and estimated (bottom) factor TACs. The results look similar as with the $\beta$-NMF for the convex case: the factorization of the 6it image is optimal for a value of $\beta$ around $0.5$, which is in agreement with the expected Poisson-Gamma nature of the noise before post-filtering. Factor modeling with $\beta=0.5$ is about $5$dB better than the one obtained from the usual Euclidean divergence relying on Gaussian noise ($\beta=2$).  In the non-convex case, due to a high dependence on the initialization, $\beta$-LMM exhibits a behavior different from the convex case. In particular, the estimated models seem to be affected by a smaller variance. This may result from the fact that the minimization algorithm likely converges to the same critical point. Indeed, for all 64 samples, the factors and factors proportions have been initialized in the same systematic way, using VCA and SUnSAL respectively. Again, the $\beta$ parameter has less impact for the 50it image which has been strongly filtered, but the optimal $\beta$ is still around 1 in the convex case. The overall performance reached in the 50it seems to be consistently better for the non-convex setting when $\beta \geq 0.5$, which may be explained by a good initialization and the joint estimation of the factor TACs.

For the 50it image, once again it is possible to see a more Poisson-like distributed noise with a higher PSNR around $30$dB with $\beta=1$. In this setting, the difference between the highest PSNR and the lowest one for $\beta=0$ is of more than $3$dB. The highest PSNR for the non-convex case is reached with $\beta=1$ and is of $32$dB.  The highest PSNR is $9$dB greater than the lowest one obtained with $\beta=0$ when estimating both TAC factors and proportions. However, the difference between the PSNR reached with $\beta=1$ and $\beta=2$ is of less than $0.5$dB. All remarks previously made for $\beta$-NMF in this case are confirmed with the results of $\beta$-LMM.

\subsection{Results on Phantom II}
\label{subsec:results_phantomII}

This paragraph discusses the results of $\beta$-SLMM obtained on Phantom II. This experiment considers both the reconstruction error (in terms of PSNR) and the estimation error for each latent variable (in terms of NMSE). The factorization with $\beta$-SLMM requires the tuning of parameter $\lambda$, which controls the sparsity of the internal variability. In this work, the value of this parameter has been empirically tuned to obtain the best possible {PSNR} result for the different values of $\beta$ and for the two 6it and 50it images. { \textit{A priori} knowledge on the binding region could also be used to adjust $\lambda$, monitoring the accuracy of the method with respect to quantitative analysis. The optimal value can thus depend on the objective of the subsequent analysis.} Two settings have been considered. In the first one, the factor TACs are fixed to their ground-truth value. Thus, the algorithm described in Section \ref{sec:bcd} updates only the factor proportions and the internal proportions $\mathbf{B}$. In this case, the factor proportions have been randomly initialized. In the second setting, the algorithm estimates the factor TACs and proportions, as well as the internal variability. In this setting, the factor proportions have been initialized using SUnSAL.

Table \ref{table:param} reports the values of $\lambda$ for each value of $\beta$ and each image. The parameters were the same for fixed and estimated $\mathbf{M}$.

\begin{table}                                                                          \renewcommand{\tabcolsep}{2pt}
\centering
\caption{Stopping criterion and variability penalization parameters.}
\label{tab:param}
\begin{tabular}{|c|c|c|c|c|c|c|}
\cline{2-6}

\multicolumn{1}{c|}{}&
\multicolumn{3}{c|}{$\lambda$}&\multicolumn{2}{c|}{$\varepsilon$}\\
\cline{2-6}
\multicolumn{1}{c|}{}&  $\beta$=0  &  $\beta$=1 &  $\beta$=2  & $\mathbf{M}$ fixed  & $\mathbf{M}$ estimated \\
\hline
6it& $1.3 \times 10^{-4}$ & $1.3 \times 10^{-3}$ &$3.9 \times 10^{-3}$  & $10^{-5}$ & $10^{-4}$  \\
\hline
50it& $6.8 \times10^{-5}$ & $6.8 \times 10^{-4}$ & $2  \times 10^{-3}$ &$10^{-5}$&$10^{-4}$ \\
\hline
\end{tabular}
\label{table:param}
\end{table}

Table \ref{table:mean_error_fixedM} presents the mean NMSE for $\mathbf{A}_1$, $\mathbf{A}_{2:K}$ and $\mathbf{A}_1 \cdot\mathbf{B}$ as well as the PSNR for the 6it and 50it images in the framework where $\mathbf{M}$ is fixed. The estimation performance of $\mathbf{A}_1 \cdot\mathbf{B}$ rather than $\mathbf{B}$ is evaluated because the partial volume effect (due to the PSF) can be propagated either in variable $\mathbf{A}_1$ or in $\mathbf{B}$. Both 6it and 50it images present similar results, with the smallest NMSE of $\mathbf{A}_1$ and $\mathbf{A}_{2:K}$ obtained for $\beta=1$ and the best estimation performance of $\mathbf{A}_1 \cdot\mathbf{B}$ obtained for $\beta=0$. However, the PSNR values show that, while 6it reaches its best performance for $\beta=0$ closely followed by $\beta=1$, 50it achieves its highest PSNR for $\beta=1$, followed by $\beta=2$. This result confirms the previous results on Phantom I, which exhibited a Poisson-Gamma noise distribution for the 6it image and a Poisson-Gaussian noise distribution for the 50it images.

\setlength{\tabcolsep}{4pt}
\begin{table}
\centering
\caption{Mean NMSE of $\mathbf{A}_1$, $\mathbf{A}_{2:K}$ and $\mathbf{A}_1 \cdot\mathbf{B}$ and PSNR of reassembled image estimated by $\beta$-LMM and $\beta$-SLMM with fixed $\mathbf{M}$ over the 64 samples, for different values of $\beta$.}
\begin{tabular}{|c|c|c|c|c|c|c|c|}
\cline{3-8}
\multicolumn{2}{c|}{}&
\multicolumn{3}{c|}{$\beta$-LMM}&\multicolumn{3}{c|}{$\beta$-SLMM}\\
\hline
\multicolumn{2}{|c|}{$\beta$} & 0  & 1  & 2 & 0  & 1  & 2 \\
\hline
\multirow{3}{*}{\rotatebox{90}{6it}}&$\mathbf{A}_1$ & 0.500 & 0.497 & \textbf{0.491} & 0.273 & \textbf{0.262} & 0.274 \\
&$\mathbf{A}_{2:K}$ & 0.304 & \textbf{0.282} & 0.290 & 0.292 & \textbf{0.267} & 0.276 \\
&$\mathbf{A}_1 \cdot \mathbf{B}$ & - & - & - & \textbf{0.423} & 0.439 & 0.492 \\

&PSNR & 28.325 & \textbf{28.345} & 28.224 & \textbf{31.905} & 31.693 & 29.825 \\
\hline
\multirow{3}{*}{\rotatebox{90}{50it}}&$\mathbf{A}_1$ & \textbf{0.447} & 0.453 & 0.452 & 0.209 & \textbf{0.196} & 0.204 \\
&$\mathbf{A}_{2:K}$ & 0.262 & \textbf{0.251} & 0.268 & 0.255 & \textbf{0.236} & 0.258 \\
&$\mathbf{A}_1 \cdot \mathbf{B}$ & - & - & - & \textbf{0.293} & 0.305 & 0.371 \\

&PSNR & 31.992 & \textbf{32.799} & 32.180 & 34.556 & \textbf{36.385} & 35.178 \\
\hline
\end{tabular}
\label{table:mean_error_fixedM}
\end{table}

Table \ref{table:mean_error} shows the mean NMSE for $\mathbf{A}_1$, $\mathbf{A}_{2:K}$, $\mathbf{\tilde{M}}^1$, $\mathbf{M}^{2:K}$ and $\mathbf{A}_1 \cdot \mathbf{B}$ in the setting where $\mathbf{M}$ is now estimated with the other latent variables. Unlike the previous experiments, the results here are less clear since, depending on the variable, different values of $\beta$ lead to the best results. This could be explained by the strong non-convexity of the problem, and possibly identifiability issues since 3 sets of latent variables need to be estimated.
The results in Table \ref{table:mean_error} show that $\beta$-LMM with $\beta=2$ performs the best for the estimation of $\mathbf{A}_{2:K}$ and $\mathbf{M}_{2:K}$ in the 6it image, and for the estimation of $\mathbf{A}_{2:K}$ in the 50it image. All variables related to specific binding, i.e., $\mathbf{A}_1$, $\mathbf{\tilde{M}}^1$ and $\mathbf{A}_1 \cdot \mathbf{B}$, are best estimated by $\beta$-SLMM with $\beta=1$. For 50it, due to the high level of filtering along with the non-convexity of this setting, analyzing the results is more difficult. It is, however, possible to state that a rather Poisson-Gaussian distributed noise yields the overall best mean NMSE of each variable.

Regarding the PSNRs, once again, the best PSNR on the 6it image is reached for $\beta=0$, closely followed by $\beta=1$. Conversely, on the 50it image, the best performance is reached for $\beta=1$, then followed by $\beta=0$. As also stated in the non-convex case of Phantom I, the initialization plays a relevant role when several sets of variables are to be estimated. This explains the differences found for the results with $\mathbf{M}$ fixed and estimated. Indeed, the high non-convexity of the problem with estimated $\mathbf{M}$ may sometimes alter the expected response.

\setlength{\tabcolsep}{4pt}
\renewcommand{\arraystretch}{1}
\begin{table}
\centering
\caption{Mean NMSE of $\mathbf{A}_1$, $\mathbf{A}_{2:K}$, $\mathbf{\tilde{M}}^1$, $\mathbf{M}^{2:K}$ and $\mathbf{A}_1 \cdot \mathbf{B}$ and PSNR of reassembled image estimated by $\beta$-LMM and $\beta$-SLMM with $\mathbf{M}$ estimated  over the 64 samples, for different values of $\beta$.}
\begin{tabular}{|c|c|c|c|c|c|c|c|}
\cline{3-8}
\multicolumn{2}{c|}{}&
\multicolumn{3}{c|}{$\beta$-LMM}&\multicolumn{3}{c|}{$\beta$-SLMM}\\
\hline
\multicolumn{2}{|c|}{$\beta$} & 0  & 1  & 2 & 0  & 1  & 2 \\
\hline
\multirow{5}{*}{\rotatebox{90}{6it}}&$\mathbf{A}_1$ & 0.382 & 0.336 & \textbf{0.327} & 0.323 & \textbf{0.311} & 0.313 \\
&$\mathbf{A}_{2:K}$ & 0.629 & 0.616 & \textbf{0.608} & 0.634 & 0.629 & \textbf{0.628} \\
&$\mathbf{\tilde{M}}^1$ & \textbf{0.300} & 0.343 & 0.375 & 0.007 & \textbf{0.006} & 0.010\\
&$\mathbf{M}^{2:K}$ & 0.356 & 0.346 & \textbf{0.306} & 0.398 & 0.390 & \textbf{0.380} \\
&$\mathbf{A}_1 \cdot \mathbf{B}$ & - & - & - & 0.475 & \textbf{0.450} & 0.686 \\
&PSNR & 27.046 & 29.445 & \textbf{30.231} & \textbf{31.301} & 30.279 & 27.178 \\
\hline
\multirow{5}{*}{\rotatebox{90}{50it}}&$\mathbf{A}_1$ & 0.482 & 0.491 & \textbf{0.472} & 0.441 & \textbf{0.423} & 0.428 \\
&$\mathbf{A}_{2:K}$ & 1.018 & 0.842 & \textbf{0.799} & 1.055 & 0.886 & \textbf{0.808} \\
&$\mathbf{\tilde{M}}^1$ & 0.430 & \textbf{0.294} & 0.332 & 0.006 & 0.004 & \textbf{0.003} \\
&$\mathbf{M}^{2:K}$ & \textbf{0.716} & 0.896 & 0.832 & \textbf{0.707} & 0.811 & 1.169 \\
&$\mathbf{A}_1 \cdot \mathbf{B}$ & - & - & - & 0.382 & 0.307 & \textbf{0.223} \\
&PSNR & \textbf{31.302} & 27.335 & 28.891 & 31.599 & \textbf{31.775} & 31.080 \\
\hline
\end{tabular}
\label{table:mean_error}
\end{table}
{Finally note that, in practice, each of the three different models evaluated above can be of interest. The most adapted model depends on the data and the application. NMF and LMM are simpler, thus less sensitive to initialization and optimization issues. On the other hand, SLMM is based on a finer modelling, and is expected to better explain the data when the specific binding factor presents some variability.}
\section{Experiments with real data}
\label{sec:real_sim}

\subsection{Real data acquisition}
To enrich our study on the impact of $\beta$ for different PET image generation settings, the experiments on real data were conducted with both a different tracer and a different scanner. More precisely, a real dynamic PET image of a stroke subject injected with [18F]DPA-714 was used to evaluate the behavior of $\beta$-SLMM in a real setting. The [18F]DPA-714 is a ligand of the 18-kDa translocator protein (TSPO) and has shown its relevance as a biomarker of neuroinflammation \cite{chauveau2009comparative}. The image of interest was acquired seven days after the stroke with an Ingenuity TF64 Tomograph from Philips Medical Systems.  The image was reconstructed  using the Blob-OS-TF algorithm \cite{Matej1996} with 3 iterations, 33 subsets and an additional postfiltering step. It consists of $L=31$ frames with durations that ranged from $10$ seconds to $5$ minutes over a total of $59$ minutes. Each frame is composed of $128 \times 128\times 90$ voxels of size $2 \times 2 \times 2$ mm$^3$. Each voxel TAC was assumed to be a mixture of $K=4$ types of elementary TACs: specific binding associated with neuroinflammation, blood, non-specific gray matter and white matter. A supervised segmentation from a registered MRI image provided a ground-truth of the stroke region, containing specific binding. The variability descriptors $\mathbf{V}$ were learned by PCA from this ground-truth. The cerebrospinal fluid was segmented and masked as a $5$th class of a K-means clustering that also provided the initialization of the factors. Factor proportions were initialized with the clustering labels found by K-means. For $\beta$-SLMM, the nominal SBF was fixed as the empirical average of TACs from the stroke region with area-under-the-curve (AUC) between the 5th and 10th percentile. Note that the reconstruction settings typically used on the Ingenuity TF64 tomograph for this kind of imaging protocol produce PET images that are characterized by a relatively high level of smoothness, inducing spatial noise correlation.

\subsection{Results}

\begin{figure}[htbp]
\includegraphics[width=0.95\columnwidth]{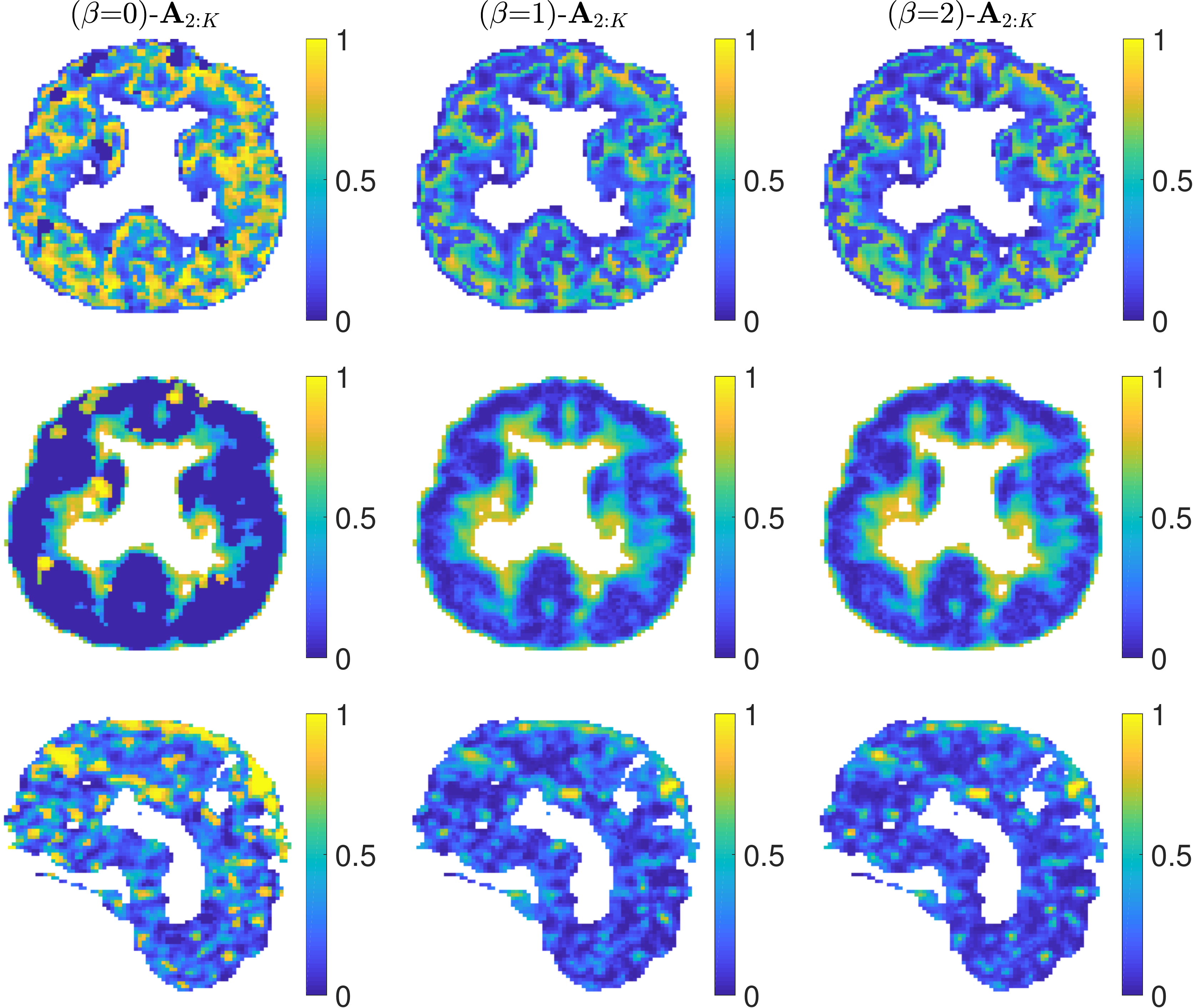}
\caption{From top do bottom: factor proportions ($\mathbf{A}_{2:K}$) from non-specific gray matter, white matter and blood obtained with $\beta$-SLMM for $\beta=0,1,2$.}
\label{fig:A_real}
\end{figure}

Figure \ref{fig:A_real} shows, from top to bottom, the factor proportions for gray matter, white matter and blood estimated by $\beta$-SLMM for $\beta\in\left\{0,1,2\right\}$ where the stopping criterion $\varepsilon$ was defined as $5 \times 10^{-4}$ and the hyperparameter $\lambda$ was set to $9 \times 10^{-2}$. { In particular, $\lambda$ was ajusted by searching for a reasonable trade-off between localization/sparsity and intensity of the variability in relevant brain areas, in particular in the central region that corresponds to the thalamus, which is also expected to be affected by the variability. Another possible strategy for choosing $\lambda$ in a clinical context would be to incorporate arterial sampling for the acquisitions of the first few patients of a given protocol.} Visual analysis suggests that all the algorithms provide a good estimation of both gray and white matters. The results for $\beta=1$ and $\beta=2$ are very similar and it is difficult to state which one achieves the best performance. This is in agreement with the synthetic results previously presented, that showed very similar estimation errors in  case of more post-reconstruction filtering. The result for $\beta=0$ is quite different from the others with more contrasted factor proportions. The sagittal view of the blood in the $3$rd row has been taken from the center of the brain. The proposed approach correctly identifies the superior sagittal sinus vein of the brain for all tested $\beta$ values. However, some clear differences can be observed and the blood is also more easily identified for $\beta=0$ than for the other values of $\beta$.

\begin{figure}[htbp]
\includegraphics[width=0.95\columnwidth]{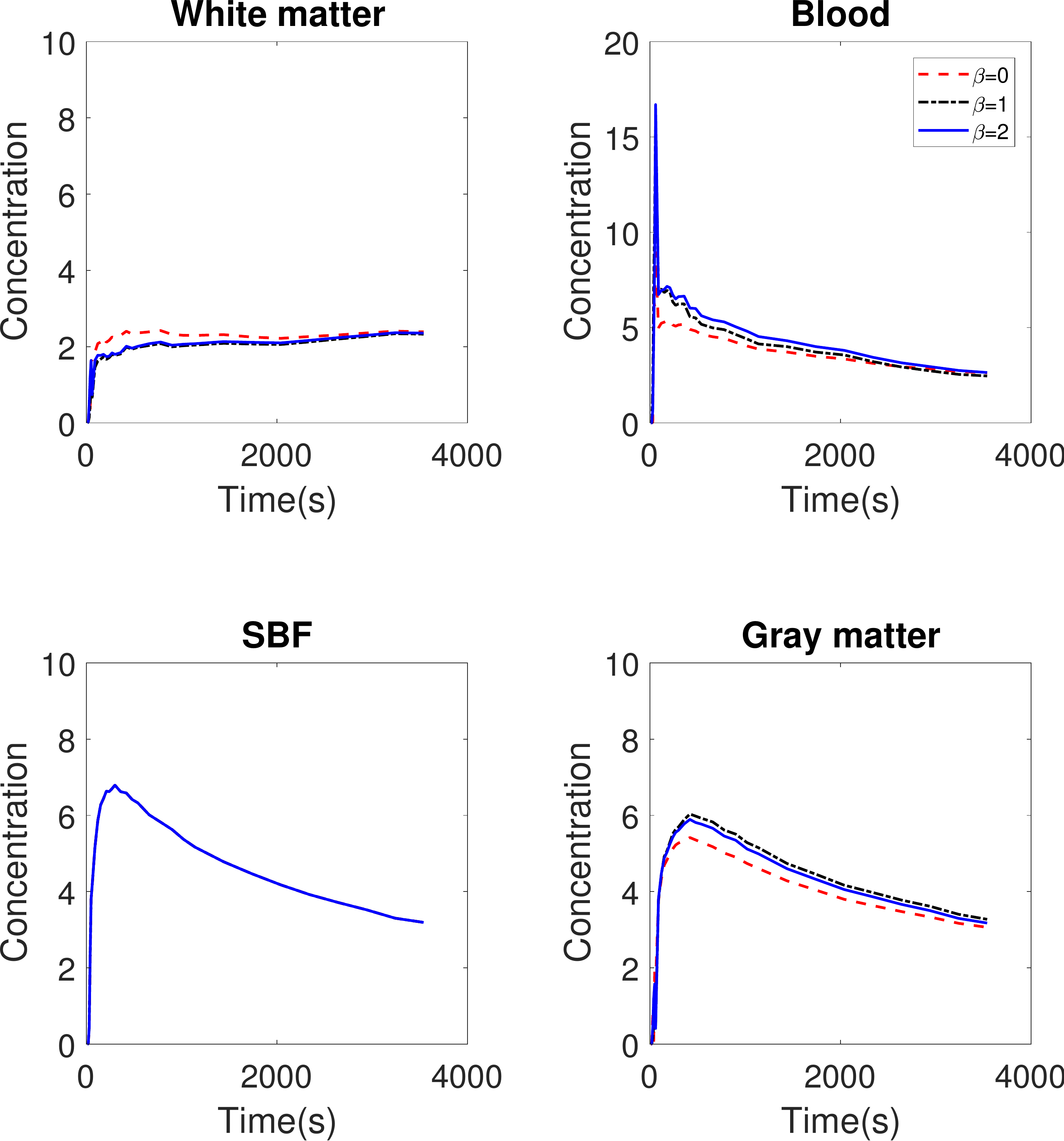}
\caption{TACs corresponding to the specific binding factor, gray matter, white matter and blood.}
\label{fig:M_real}
\end{figure}

Figure \ref{fig:M_real} confirms these findings, showing TACs that are very similar for $\beta\in \left\{1,2\right\}$ while the TACs for $\beta=0$ are always a bit apart from the others. The expected initial pick characterizing the blood TAC is more easily identified with $\beta=1$ and $\beta=2$. On the other hand, for $\beta=0$ the TAC associated with the non-specific gray matter has a lower AUC than the two others, further differentiating from the specific binding TAC.

\begin{figure*}[htbp]
\includegraphics[width=0.95\textwidth]{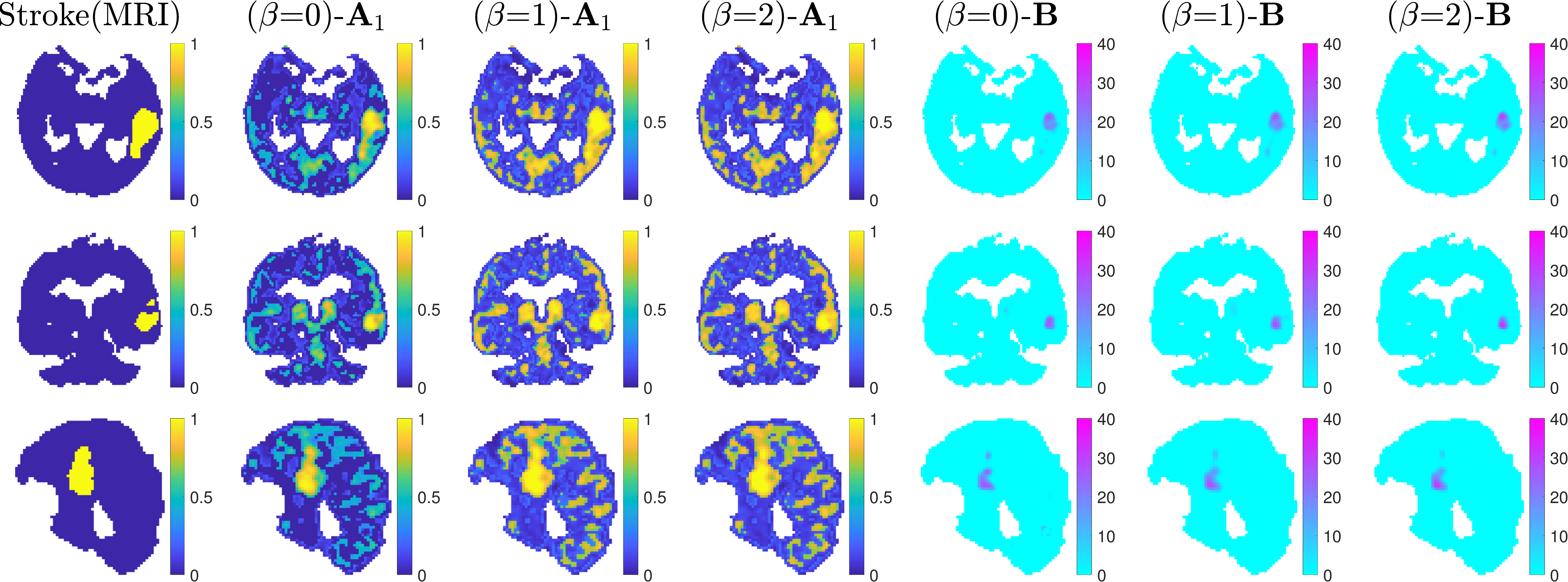}
\caption{From left to right: {Transversal, coronal and sagital planes (top to bottom) of }MRI ground truth of the stroke zone, factor proportions ($\mathbf{A}_{1}$) from specific gray matter and variability matrices ($\mathbf{B}$) obtained with $\beta$-SLMM for $\beta=0,1,2$.}
\label{fig:3D_SBR_real}
\end{figure*}

Figure \ref{fig:3D_SBR_real} shows a manually segmented ground-truth of the stroke zone along with the corresponding factor proportions and variability matrices estimated with SLMM. The results obtained with $\beta=0$ show a more accurate identification of the stroke zone. Results with $\beta=1,2$ are very similar: they localize the thalamus, known for having higher binding of neuroinflammation. But they also recover non-specific gray matter in the factor proportion related to specific binding. All values of $\beta$ show variability matrices that are consistent with the stroke area.

The results for $\beta\in\left\{1,2\right\}$ are very similar but $\beta=2$ shows a stronger intensity, while $\beta=1$ shows a more spread result, even presenting the influence of the thalamus in the $2$nd row, similarly to $\beta=0$.

\section{ Discussion}
\label{sec:discuss}
\begin{figure}
    	\centering
		\includegraphics[width=0.48\linewidth]{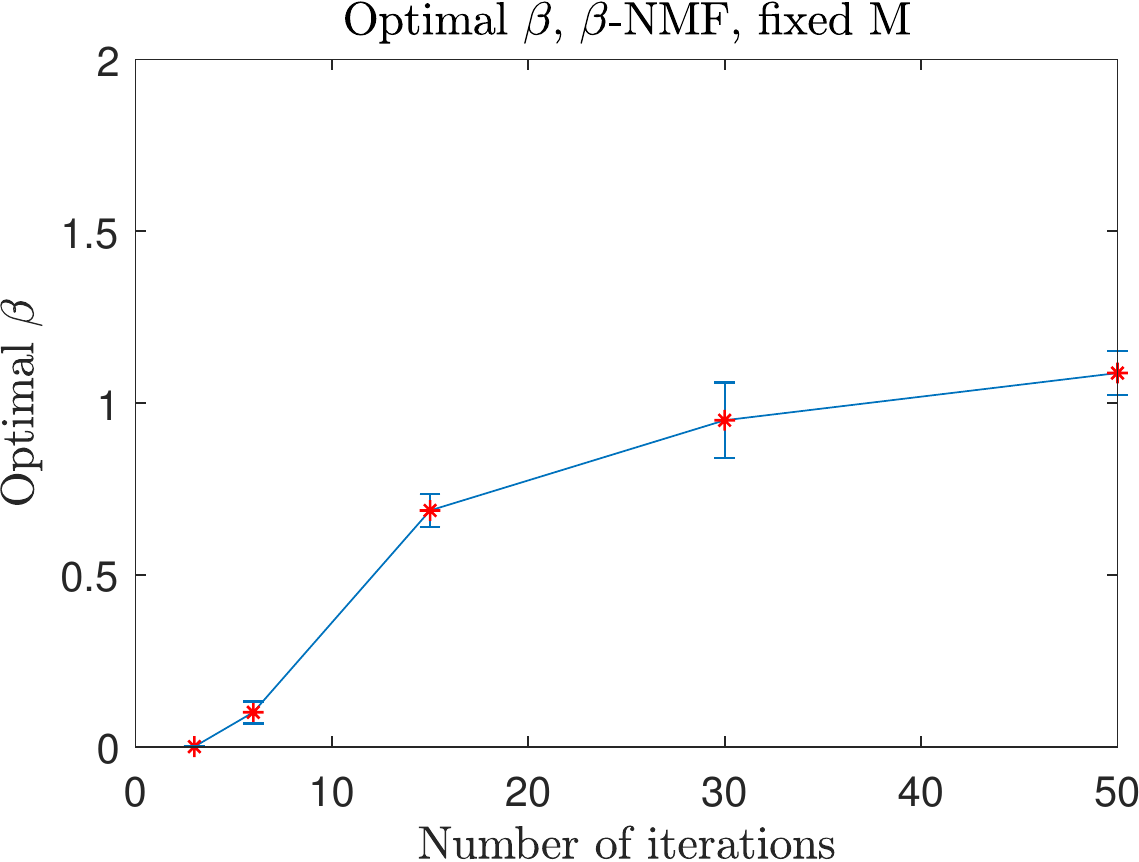}
		\includegraphics[width=0.48\linewidth]{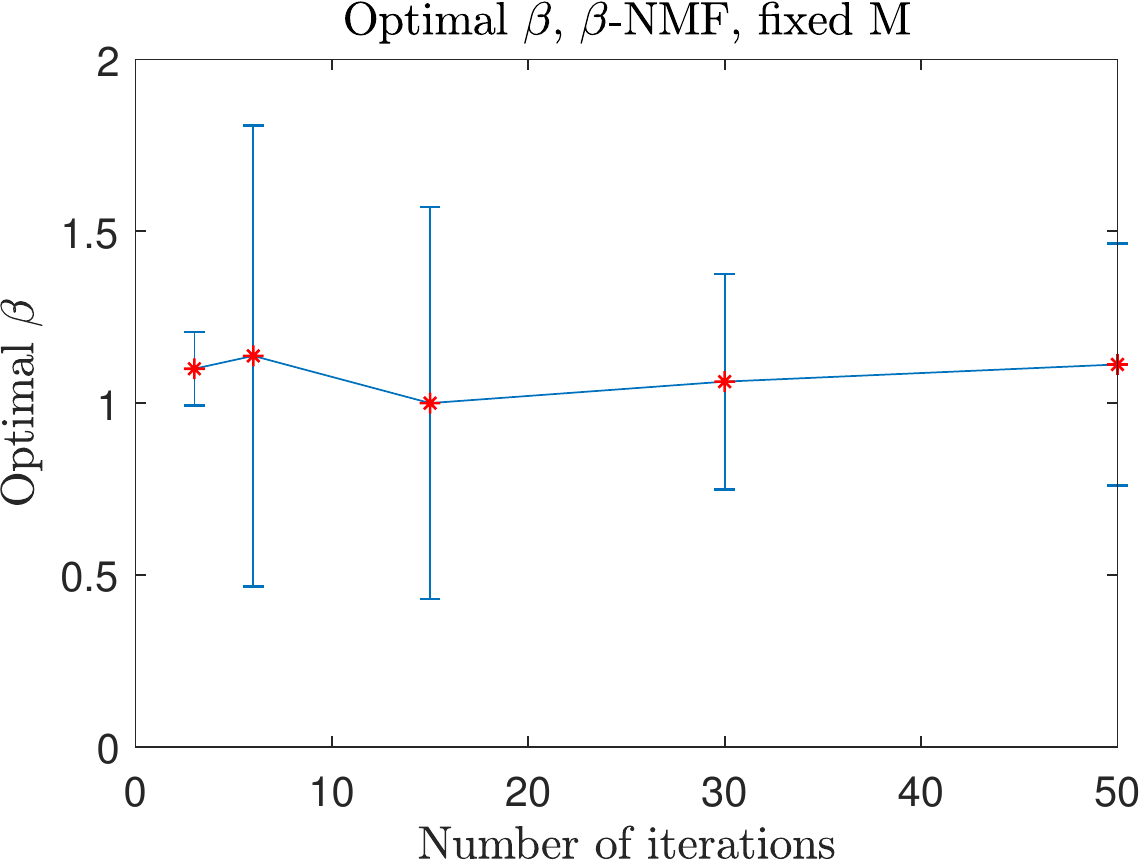}
			\caption{Optimal beta  computed with the $\beta$-NMF algorithm in the convex setting for 3it, 6it, 15it, 30it and 50it over:  (left)  16 samples without a postfiltering step,  (right)  16 samples with a postfiltering step.}%
            \label{fig:optbeta}%
	\end{figure}

{
As previously discussed, different acquisition conditions and reconstruction settings produce PET images with different noise distributions. Therefore, the optimal value of $\beta$, i.e. the value which produces the best decomposition, highly depends on the experimental setting. This can be observed in the above-presented experiments, where the optimal $\beta$ was shown to be driven by the reconstruction, the model, and even the way we evaluate the factor decomposition.

One of the main objectives of this paper was to demonstrate the flexibility of the $\beta$-divergence, and its ability to improve the factor analysis even when the noise is not well characterized. However, this can also be seen as a weakness, because how to choose $\beta$ in real situations is not straightforward. As a tentative to address this issue, we studied the optimal $\beta$ value for synthetic images generated with the same process described in paragraph \ref{subsec:synth_sim_gen} for 3, 6, 15, 30 and 50 reconstruction iterations (respectively 3it, 6it, 15it, 30it and 50it images). We run 16 independent simulations for each setting, and evaluated the optimal $\beta$ as a function of reconstruction iterations, with and without final post-filtering. Figure \ref{fig:optbeta} shows the  optimal $\beta$ for 3it, 6it, 15it, 30it and 50it, computed over 16 samples without a postfiltering step (left) and with a postfiltering step (right). This figure can serve as a reference to choose $\beta$ in this experimental setting, and it is consistent with the other results presented above. To summarize, without the post-filtering step, a reasonable choice of $\beta$ is around $0.5$ for few iterations, and $1$ or slightly above for more iterations. We also remark that the influence of $\beta$ is less clear when a post-filtering step has been used within reconstruction.

This strategy is expected to remain valid for other tracers, other cameras or other reconstruction algorithms. Specific numerical simulations dedicated to the experimental setting can be conducted to obtain a relevant tuning of the $\beta$.

Moreover, throughout this article, the main measure of evaluation was the PSNR. A more insightful evaluation could be obtained by separately measuring the final bias and variance for each setting. To further enlighten the interest of using a correct data-fitting measure, this study was conducted on Phantom I. The analysis showed that the bias is the most relevant element for the final PSNR, i.e., it is the measure that is most affected by the use of different values of $\beta$ (see \cite{Cavalcanti2018TR} for more details).
}

\section{Conclusion}
\label{sec:concl}

This paper studied the role of the data-fidelity term when conducting factor analysis of dynamic PET images. We focused on the beta-divergence, for which the NMF and LMM decompositions were already proposed in other applicative contexts. We also introduced a new algorithm for computing a factor analysis allowing for variable specific-binding factor, termed $\beta$-SLMM.

For all those three models, experimental results showed the interest of using the $\beta$-divergence in place of the standard least-square distance. The factor and proportion estimations were indeed more accurate when computed with an suitable value of $\beta$. The improvement was shown to be higher when the image had not suffered too strong post-processing corrections. The $\beta$-divergence thus appeared to be a general and flexible framework for analyzing different kind of dynamic PET images.

Future works should consider the use of the $\beta$-divergence in the whole image processing pipeline, including the reconstruction from the sinograms and the denoising. This should further improve the final factor analysis results. While the scope of this paper was to study the relevance of a flexible divergence measure in PET image processing, a deeper evaluation of the impact of the method on input function estimation and quantification parameter estimation  within clinical applications for which arterial sampling is available should also be envisaged in the future.

\bibliographystyle{ieeetran_shorten} 
\bibliography{strings_all_ref,bibli}

\end{document}